\documentclass[12pt,preprint,a4paper]{article}

\usepackage{amsmath, amssymb, mathrsfs, graphicx, subcaption, dsfont, enumerate, epstopdf, xcolor}
\usepackage{cite} 
\usepackage[colorlinks, allcolors=blue!70!black, linktocpage]{hyperref}
\usepackage[inline]{enumitem}
\usepackage[utf8]{inputenc}

\numberwithin{equation}{section}

\def\half{\frac{1}{2}}
\def\eq#1 { \begin{equation} #1 \end{equation} }

\def\a{\alpha}

\def\e{\epsilon}
\def\w{\omega}
\def\k{\kappa}

\def\d{\partial}
\def\cL{\mathcal{L}}

\def\cA{\mathcal{A}}

\def\cF{\mathcal{F}}

\def\cO{\mathcal{O}}

\def\cB{\mathcal{B}}
\def\cS{\mathcal{S}}

\def\sl2r{SL(2,\mathbb{R})}

\def\ce{\varepsilon}

\newcommand{\lsim}{\mathrel{\hbox{\rlap{\lower.55ex \hbox{$\sim$}} \kern-.3em \raise.4ex \hbox{$<$}}}}
\newcommand{\gsim}{\mathrel{\hbox{\rlap{\lower.55ex \hbox{$\sim$}} \kern-.3em \raise.4ex \hbox{$>$}}}}

 \newcommand{\be}{\begin{equation}}
\newcommand{\ee}{\end{equation}}

\newcommand{\vev}[1]{\left< #1 \right>} 

\usepackage{bm}

\newcommand{\lb}{\left}
\newcommand{\rb}{\right}

\newcommand{\mc}{\mathcal}

\newcommand{\df}[1]{\boldsymbol{#1}}

\makeatletter
\newcommand{\strong}[1]{\@strong{#1}}
\newcommand{\@@strong}[1]{\textbf{\let\@strong\@@@strong#1}}
\newcommand{\@@@strong}[1]{\textnormal{\let\@strong\@@strong#1}}
\let\@strong\@@strong
\makeatother

\begin{document}

\title{\begin{flushright}\vspace{-1in}
       \mbox{\normalsize  EFI-16-7}
       \end{flushright}
       \vskip 20pt
Covariant effective action for a Galilean invariant quantum Hall system}

\date{\today}

\author{
	Michael Geracie\thanks{\href{mailto:mgeracie@uchicago.edu}         {mgeracie@uchicago.edu}} ,
	Kartik Prabhu\thanks{\href{mailto:kartikp@uchicago.edu}         {kartikp@uchicago.edu}}  ~and 
	Matthew M. Roberts\thanks{\href{mailto:matthewroberts@uchicago.edu}
{matthewroberts@uchicago.edu}}~
      \\[15pt]
   {\it Kadanoff Center for Theoretical Physics,}\\
   {\it Enrico Fermi Institute and Department of Physics}\\
   {\it University of Chicago, Chicago, IL 60637 USA}
} 

\maketitle

\begin{abstract}
We construct effective field theories for gapped quantum Hall systems coupled to background geometries with local Galilean invariance i.e. Bargmann spacetimes. Along with an electromagnetic field, these backgrounds include the effects of curved Galilean spacetimes, including torsion and a gravitational field, allowing us to study charge, energy, stress and mass currents within a unified framework. A shift symmetry specific to single constituent theories constraints the effective action to couple to an effective background gauge field and spin connection that is solved for by a self-consistent equation, providing a manifestly covariant extension of Hoyos and Son's improvement terms to arbitrary order in $m$.
\end{abstract}

\newpage
\tableofcontents

\section{Introduction}\label{sec:intro}

The salient property of quantum Hall systems is that their DC Hall conductance \(\sigma_{xy}\) and the charge of the state \(Q\) are given by simple functions of the filling fraction \(\nu\) and the magnetic flux \(N_\phi = B/2\pi\)
\eq{
\sigma_{xy} = \frac{\nu}{2\pi},  \qquad \qquad Q = \nu N_\phi = \frac{\nu}{2\pi}B.
}
where \(B\) is the magnetic field. It is well known that these properties are accounted for by a Chern-Simons term in the effective action
\eq{
S_{CS} = \frac{\nu}{4\pi} \ce^{\mu\nu\rho}A_\mu \d_\nu A_\rho.
}

 It is also known that quantum Hall systems exhibit a second topological property, the \emph{shift} \cite{Wen:1992ej}. This new topological quantity is related to a non-dissipative transport coefficient, the Hall viscosity \cite{Avron:1995fg, Avron:1997, Read:2008rn, ReadRezayi:2010}. For a quantum Hall state on a sphere with $N_\phi$ units of magnetic flux the shift \(\cS\) is an offset in the charge of the state,
\eq{
Q({\rm sphere}) = \nu (N_\phi + \cS).
}
More generally on a Riemann surface of genus $g$, the total charge is
\eq{\label{eq:charge-shift}
Q(\Sigma_g) = \nu (N_\phi + (1-g)\cS).
}
Wen and Zee \cite{Wen:1992ej} reproduced this by adding a mixed Chern-Simons term to the effective action,
\eq{
S_{WZ} = \frac{\k}{2\pi} \ce^{\mu\nu\rho}\w_\mu \d_\nu A_\rho \qquad \qquad \text{where}\qquad \qquad\kappa = \half \nu \cS,
}
and $\w_\mu$ is the \(SO(2)\)-spin connection for local spatial rotations. This term also predicts a non-dissipative viscosity coefficient \(\eta_H\) called the \emph{Hall viscosity}
\eq{
T^{ij} = \eta_H \ce^{k(i}\delta^{j)\ell} \d_t{\delta h_{k\ell}} \qquad \qquad\text{with}\qquad \qquad \eta_H = \frac{\kappa B}{4\pi}.\label{eq:HS_viscosity_prediction}
}

The bulk transport in such topological phases of matter was studied in \cite{Bradlyn:2014wla,Gromov:2014vla} without assuming Galilean (or Lorentz) invariance. One would expect Galilean invariance to place further restrictions on the effective action describing quantum Hall systems.
	The construction is more subtle systems with Galilean boost symmetry, $t\rightarrow t, x \rightarrow x - vt$, which is a symmetry of non-relativistic electrons interacting through Coulomb interactions.\footnote{In real quantum Hall systems there is both an underlying lattice and disorder, both of which break the boost invariance of the Schr\"odinger equation. We may expect that on large enough scales the ionic lattice is screened and the disorder self-cancels, there is an approximate Galilean boost symmetry. Alternatively one may simply consider disorder as additional background data breaking the Galilean symmetry of the Schr\"odinger equation.} The key problem is that a single constituent system cannot distinguish between electromagnetic and gravitational potentials, and hence they must appear together in an effective action. Unfortunately, while the electromagnetic potential is invariant under Galilean boosts, the gravitational potential is not (see \autoref{sec:barg-geometry}) and so simply replacing the electromagnetic potential by their combination yields an action which is not boost invariant.\\

A manifestly covariant way to describe Galilean invariant systems is using Newton-Cartan geometry, developed initially to obtain a covariant description of Newtonian gravity in the spirit of General Relativity (see Ch.12 of \cite{MTW-book} and Ch.4 of \cite{Mal-book}, and \cite{GPR_geometry} for a detailed set of references). An early attempt at a Galilean invariant construction of quantum Hall effective actions is the work of Hoyos and Son \cite{Hoyos:2011ez}. Using the method of non-relativistic general coordinate invariance, they constructed corrections to the Chern-Simons and Wen-Zee terms to recover Galilean invariance to a fixed order in a derivative expansion (or equivalently, an expansion in mass \(m\)). Using this method, they showed that in Galilean invariant quantum Hall states made of quasiparticles of a single charge-to-mass ratio, the shift \(\cS\) (or equivalently the Hall viscosity \(\eta_H\)) can be measured via current response to inhomogeneous electric and magnetic fields. Specifically, the currents are determined by two quantities $\kappa$ and $\e''(B)$,
\eq{
j^x =\frac{\nu}{2\pi}E^y - \left(\frac{\k}{4\pi B} - \frac{m \e''(B)}{B}\right)\nabla^2 E^y , \qquad \qquad j^i = - \e''(B) \ce^{ij} \d_j \delta B.\label{eq:HS_main_prediction}
}
and therefore these two possible measurements could allow one to experimentally determine $\kappa$ and therefore the shift.\footnote{The contribution proportional to $\kappa$ in the first equation coming from the Wen-Zee term  was also found independently by Bradlyn et.~al.~\cite{Bradlyn:2012ea}.} A similar method was used by Gromov and Abanov in \cite{Gromov:2014gta} to relate the density-curvature response to the chiral central charge.

 A careful look reveals that the description in terms of modified diffeormorphisms of the electromagnetic potential is simply a consequence of conflating the electromagnetic and gravitational potentials (see \cite{Jensen:2014aia,Jensen:2014ama,GPR_geometry, GPR-fluids} for details or \autoref{sec:barg-geometry} for an overview). Using a set of local frames and coframes (with which local observers make measurements) and a way to translate between different frames via local (possibly spacetime dependent) Galilean transformations, one can construct a description of Newton-Cartan geometry in direct analogy to the coframe/tetrad description of Riemannian and Lorentzian geometry. The structure of the Bargmann group (the group containing Galilean transformations, spatial and temporal translations, and a conserved particle number) which describes massive matter systems, dictates that the gravitational field most naturally arises as a component of an ``extended coframe". Using this, one can then construct general Bargmann geometries (including torsion) that completely describe the behavior of massive non-relativistic systems in a manifestly Galilean-covariant manner \cite{Jensen:2014aia,GPR_geometry,BM}.\\

The goal of this work is to use the covariant formulation of Bargmann geometry \cite{GPR_geometry,GPR-fluids} to describe the effective theory of quantum Hall systems. Thus, our work can be seen as a generalization of \cite{Bradlyn:2014wla} to include restrictions due to Galilean boost symmetry, or as a covariant formulation of \cite{Hoyos:2011ez} that extends their results to any order in the derivative expansion. We construct a manifestly Galilean invariant effective action describing the dynamics of a quantum Hall system in a Bargmann spacetime. We show that this recovers the results (\ref{eq:HS_viscosity_prediction},\ref{eq:HS_main_prediction}) of \cite{Hoyos:2011ez}. We also calculate response to homogeneous electric fields to confirm the Kohn-Luttinger theorem.  Since the quantum Hall system can be coupled to general Bargmann spacetimes (including torsion), this approach also allows us to compute energy currents, which was not possible before. We confirm that energy currents are purely of the form of an energy magnetization, as is to be expected from a gapped system.\\

In \autoref{sec:newton-cartan} we review the non-relativistic Bargmann geometry constructed in \cite{GPR_geometry} and the definitions of matter currents. We recall the expected symmetries of a single component microscopic description of the quantum Hall effect in \autoref{sec:symmetries}. In \autoref{sec:frame} we define an effective drift velocity due to electromagnetic and gravitational effects for the quantum Hall system which is manifestly invariant under the symmetries of the microscopic description, and use it to construct the effective gauge field and Galilean connection that couple to the Hall system in the effective theory. We propose a manifestly invariant effective action for the quantum Hall system in \autoref{sec:eff-action}, and compare our approach to that of Hoyos and Son (the explicit computations for the comparison are collected in \autoref{app:HoyosSon}). We collect our results on the linear response, including the energy currents in \autoref{sec:lin-response}.\\

We will use an abstract index notation for tensors throughout this paper. We use indices \(\mu,\nu,\lambda,\ldots\) denote tensors on spacetime while the indices \(i,j,k,\ldots\) denote the spatial coordinate components in a local coordinate system \((t,x^i)\). The vector representation of the Galilean group will carry indices \(A,B,C,\ldots\) while the representation of spatial rotations will have indices \(a,b,c,\ldots\). The ``extended representation" of the Galilean group will be denoted by \(I,J,K,\ldots\). We denote differential forms by a boldface letter when using an index free notation. Our sign and factor conventions follow those of \cite{Wald-book}.

\section{Overview of non-relativistic geometry}\label{sec:newton-cartan}

Let us begin with a brief review of Newton-Cartan and Bargmann geometry, the natural setting for non-relativistic physics. It will be convenient to formulate the geometry in terms of coframes or vielbeins. Readers familiar with this formulation, and specifically the conventions of \cite{GPR-fluids,GPR_geometry}, should feel free to skip to \autoref{sec:symmetries}.

\subsection{Frame formulation of Newton-Cartan geometry}\label{firstOrder}
Observers perform measurements with respect to a local set of frames $e^\mu_A$ and coframes $e^A_\mu$, $A = 0 , 1 , \dots d$ (here spacetime is $d+1$ dimensional and 0 signifies the temporal direction) of linearly independent vectors and forms, which are left- and right-inverses of each other, $e^A_\mu e^\mu_B = \delta^A_B,~e^\mu_A e^A_\nu = \delta^\mu_\nu$. Since we want to consider non-relativistic physical systems we demand that different choices of local frames be related by a local Galilean transformation as
\begin{align}\label{localGal}
	\df e'^{A} = \Lambda^{A}{}_B \df e^B \qquad \qquad\text{that is}\qquad \qquad
	\begin{pmatrix}
		\df e'^{0} \\
		\df e'^{a}
	\end{pmatrix}
	=
	\begin{pmatrix}
		1 & 0 \\
		- k^{a}
		 & \Theta^{a}{}_b
	\end{pmatrix}
	\begin{pmatrix}
		\df e^0 \\
		\df e^{b}
	\end{pmatrix}.
\end{align}
where \(\Lambda^{A}{}_B\) parametrizes the Galilean group $Gal(d)$ with \(k^a\) corresponding to local Galilean boosts, and $\Theta^a{}_b \in SO(d)$ corresponding to local spatial rotations. There are two tensors which are invariant under the action of the Galilean group given by
\begin{align}\label{invariants}
	&n_A = 
	\begin{pmatrix}
		1 & 0 & \dots & 0 
	\end{pmatrix},
	&&h^{AB} =
	\begin{pmatrix}
		0 & 0 \\
		0 & \delta^{ab}
	\end{pmatrix}
\end{align}

In a flat geometry, one may choose frames aligned with an inertial coordinate system $\df e^A= d x^A$ where $x^A = ( t , x^1 , \dots x^d)$. Changing to another inertial coordinate system $t'=t, x'^a = \Theta^a{}_b x^b - k^a t$ by rotating by $\Theta^a{}_b$ and boosting by a velocity $k^a$ performs a constant Galilean transformation (\ref{localGal}).

If we wish to take derivatives in a covariant fashion, we also need a spin connection $\df\omega^A{}_B$ valued in the Lie algebra of the Galilean group. Equivalently, the \(1\)-form $\df \omega^A{}_B$ satisfies the algebraic constraints
\be\label{lieGal}
	n_A \df \omega{}^A{}_B = 0 , \qquad \qquad \df \omega^{(AB)} = 0,
\ee
and transforms as a connection under Galilean transformations
\begin{align}\label{connection}
	{\df \omega'}^{A}{}_{B} = \Lambda^{A}{}_C \df \omega^C{}_D ( \Lambda^{-1} )^D{}_{B} + \Lambda^{A}{}_C d ( \Lambda^{-1} )^C{}_{B} .
\end{align}
The conditions (\ref{lieGal}) are equivalent to $n_A$ and $h^{AB}$ being covariantly constant
\be
	D n_A = 0 , \qquad \qquad D h^{AB} = 0,
\ee
where covariant exterior derivatives are taken in the usual way on quantities with raised or lowered indices \(A,B,\ldots\), for example
\begin{align}
	D \alpha^A = d \alpha^A + \df \omega^A{}_B \alpha^B ,
	&& D \beta_{A} = d \beta_A - \beta_B \df\omega^B{}_A.
\end{align}
The transformation law (\ref{connection}) ensures that the covariant derivatives transform properly under $Gal(d)$.

What local geometry is encoded in this data? Using the invariants (\ref{invariants}) we can construct the invariant spacetime tensors
\begin{align}
	n_\mu = n_A e^A_\mu ,
	\qquad \qquad
	h^{\mu \nu} = h^{AB} e_A^\mu e_B^\nu .
\end{align}
The tensor field $h^{\mu \nu}$ is positive semi-definite and serves as a ``spatial metric". It's kernel is spanned by the non-vanishing form $n_\mu$, which serves as a ``temporal metric"\footnote{We place this terminology in quotes as neither of these are metrics in the strict sense, since they are degenerate.} sometimes called the clock form. We can also use the invariant totally antisymmetric tensors $\e^{AB\ldots C},~ \e_{AB\ldots C}$ to construct spacetime antisymmetric tensors
\eq{
\ce^{\mu\nu \ldots \lambda} = e^\mu_A \ldots e^\lambda_C \e^{A\ldots C}
}
and similarly for $\ce_{\mu\nu \ldots \lambda}$. Here we use the sign convention $\e^{01\ldots d} = \e_{01\ldots d}=+1.$

The spin connection defines a derivative operator via
\begin{align}
	\nabla_\mu e_\nu^A  = \d_\mu e^A_\nu -\Gamma^\lambda_{\mu\nu} e^A_\lambda= - \omega_\mu{}^A{}_B e^B_\nu .
\end{align}
This is a Galilean invariant definition by virtue of (\ref{localGal}) and (\ref{connection}) and the conditions (\ref{lieGal}) are equivalent to the``metric compatibility" of $\nabla$
\be\label{metricCompatibility}
	\nabla_\mu n_\nu = 0  , \qquad \qquad \nabla_\lambda h^{\mu \nu} = 0.
\ee
The data $(e^A, \omega^A{}_B)$ is then equivalent to $(n , h, \nabla)$, the language in which Newton-Cartan geometry is usually phrased in the literature.

\subsection{The gravitational potential, torsion and Bargmann geometry}\label{sec:barg-geometry}

There is however an additional background field always present in massive non-relativistic theories and which we shall get a lot of mileage out of: the Newtonian gravitational potential $V$. This also transforms under local Galilean transformations, though the reason is perhaps subtle. Consider for example a Schr\"odinger field $\psi$ of mass $m$ in flat space coupled to a gravitational field
\begin{align}
	S = \int dt d^dx \left( i \psi^\dagger \partial_t \psi - \frac 1 {2m} \partial^i \psi^\dagger \partial_i \psi - m V \psi^\dagger \psi \right) .
\end{align}
The same field in a boosted frame is given by
\begin{align}\label{schrodTransf}
	\psi'(x') = e^{ i  \frac 1 2 m k^2 t - i m k_i x^i }\psi(x)  .
\end{align}
This ensures the invariance of the Schr\"odinger equation by taking momentum eigenstates of momentum $p^i$ to eigenstates of momentum $p^i - m k^i$ (see for instance \cite{Kuchar-Sch}).

For an equivalent description, remove the phase factor in (\ref{schrodTransf}) by a redefinition of fields. In this picture $\psi'(x') = \psi (x)$, but
\begin{align}
	S = \int dt' d^dx' \left( i \psi'^\dagger \partial_{t'} \psi' - \frac 1 {2m} | ( \partial'_i - i m k_i ) \psi' |^2 - m \left( V + \frac 1 2 k^2 \right) \psi'^\dagger \psi' \right)
\end{align}
so that $\psi'$ is the wavefunction of a system in an external vector potential $a_\mu = \begin{pmatrix} - V - \frac 1 2 k^2 , & k_i \end{pmatrix}$ that couples to mass via $D_\mu = \partial_\mu - i m a_\mu$. This mass gauge field transforms under boosts as
\begin{align}\label{alphaTransf}
	a'_0 = a_0 - \frac 1 2 k^2,
	&&a'_i = a_i + k_i .
\end{align}

We've thus replaced $V$ with a vector $a_\mu$ that couples to to the $U(1)$ current corresponding to mass conservation and that transforms in a non-trivial way under boosts. Despite the additional components we've assigned to it, $a_\mu$ does not in fact contain more information than $V$, since $d$ of it's components are pure gauge. The advantage of this description is that (\ref{alphaTransf}) may be generalized to curved spacetimes while (\ref{schrodTransf}) cannot. Our presentation began with flat-space Schr\"odinger, but the boost transformation of $a_\mu$ can be seen in many ways. It was demonstrated to be equivalent to Son and Wingate's modified diffeomorphisms in \cite{Jensen:2014aia} and derived directly from a coset construction on the Bargmann group in this work and others \cite{GPR_geometry, Karananas:2016hrm}. We present the discussion above as an alternate take on what may at first seem a somewhat unnatural description of Newtonian gravity.\\

In curved spacetimes and for a general $Gal(d)$ transformation, the generalization of (\ref{alphaTransf}) turns out to be $\df a' = \df a +k_b \Theta^b{}_a \df e^a - \frac 1 2 k^2 \df n $. This formula is somewhat inconvenient to work with directly since it is nonlinear in $\df a$. However, if we group $\df a$ with the coframes $\df e^A$ into a single object, the transformation is linear and is simply a $d+2$-dimensional representation of $Gal(d)$ called the extended representation\footnote{This is in fact natural in the coset construction, where one is forced by consistency to break the generator $M$ of mass along with the generators $P_A$ of spacetime translations which together transform in the extended representation $P_I = \begin{pmatrix} P_A & M \end{pmatrix}$. The coframe and mass gauge field then come collected together as the extended coframe in the broken part of the Mauer-Cartan form $\df\omega_{MC} = \df e^I P_I + \df \w_{IJ} M^{IJ}$.}
\begin{align}\label{extendedRep}
	\begin{pmatrix}
		\df  n' \\
		\df e'^a \\
		\df a'
	\end{pmatrix}
	=
	\begin{pmatrix}
		1 & 0 & 0 \\
		- k^a & \Theta^a{}_b & 0 \\
		- \frac 1 2 k^2 & k_a \Theta^a{}_b & 1 
	\end{pmatrix}
	\begin{pmatrix}
		\df n \\
		\df e^b \\
		\df a
	\end{pmatrix}.
\end{align}
We shall refer to this object as the extended coframe $\df e^I$
\be
	\df e^I = 
	\begin{pmatrix}
		\df e^A \\
		\df a
	\end{pmatrix},
	\qquad \qquad\text{with}\qquad \qquad \df e'^I = \Lambda^I{}_J \df e^J
\ee
where $\Lambda^I{}_J$ is the matrix appearing in (\ref{extendedRep}). From here on we shall always denote $Gal(d)$ vector and covector indices by $A,B,\dots$ and extended indices by $I,J,\dots$.

The extended representation also has two invariant tensors
\be
	n_I = 
	\begin{pmatrix}
		1 & 0 & \cdots & 0 \\
	\end{pmatrix}
	, \qquad \qquad
	g^{IJ} = 
	\begin{pmatrix}
		0 & 0 & 1 \\
		0 & \delta^{ab} & 0 \\
		1 & 0 & 0 
	\end{pmatrix}.
\ee
The first is simply an extension of $n_A$ to one higher dimension, but the tensor $g^{IJ}$ is actually a metric of Lorentzian signature. We shall use it throughout to freely raise and lower extended indices. There is also an invariant projector 
\begin{align}
	\Pi^A{}_I = 
	\begin{pmatrix}
		1 & 0 & 0 \\
		0 & \delta^a{}_b & 0 
	\end{pmatrix}
\end{align}
which we use to project extended indices to the vector representation or pull back covector indices to the extended representation. For example
\be
	n_I = n_A \Pi^A{}_I,
	 \qquad \qquad h^{AB} = \Pi^A{}_I \Pi^B{}_J g^{IJ} .
\ee
It is often much easier to work with objects in the extended representation due to the existence of an invertible metric.\\

One may define the spacetime torsion in complete analogy with Riemannian geometry as
\begin{align}\label{firstStructure}
	\df T^A = D \df e^A .
\end{align}
This is known as the first Cartan structure equation and is equivalent to the definition $(\nabla_\mu \nabla_\nu - \nabla_\nu \nabla_\mu) f = - T^\lambda{}_{\mu \nu} \nabla_\lambda f$ for any function $f$.

In the Lorentzian case, the data in the torsion tensor is completely equivalent to that in the spin connection as it is simply an algebraic equation in $\df T^A$ and $\df\omega^A{}_B$. One may either consider (\ref{firstStructure}) as the definition of torsion given a connection or as determining the connection given a torsion tensor. This is not the case in the Galilean setting since one of the components of (\ref{firstStructure}) does not involve the connection at all. Contracting with $n_A$ and using the first property in (\ref{lieGal}), we have simply the constraint $\df T^0 = d\df n$. The first structure equation then does not have enough data to determine the connection given only $\df T^A$ and $\df e^A$ (there are $\half d(d+1)^2$ components of the connection but only $\half d^2(d+1)$ equations constraining them). In the metric formalism this is equivalent to the statement found throughout the Newton-Cartan literature that there are many derivative operators with a particular torsion satisfying the metric compatibility conditions (\ref{metricCompatibility}).

However, this not a deficiency of the treatment, but rather an essential feature. It is well known that unlike the relativistic case, Newtonian gravitational effects are not contained in metric data, but in the part of the derivative operator that is undetermined by them (see \cite{MTW-book,Mal-book, GPR_geometry}). On the other hand, we have claimed above that Newtonian gravity is encoded in the final component of an extended coframe $\df e^I$. Let us see how we can retrieve the description found in \cite{MTW-book,Mal-book} from this fact.

Just like the extended coframes \(\df e^I\) we have the extended representation $\df\omega^I{}_J$ of the Galilean connection
\be\label{lieRep}
	\df \omega^I{}_J = 
	\begin{pmatrix}
		0 & 0 & 0 \\
		\df \varpi^a & \df \omega^a{}_b & 0 \\
		0 & - \df \varpi_b & 0 
	\end{pmatrix}.
\ee
Using this we can define the extended torsion tensor (see \cite{GPR_geometry})
\be\label{firstStructureExtended}
	\df T^I = D\df e^I = d \df e^I + \df \omega^I{}_J \wedge \df e^J .
\ee
which in components reads
\be\label{eq:T-decomp}
	\df T^I = 
	\begin{pmatrix}
		n_I \df T^I\\ \df T^a \\ \df f
	\end{pmatrix} =
	\begin{pmatrix}
		d\df n \\ d\df e^a + {\df \omega^a}_b \wedge \df e^b + \df\varpi^a \wedge \df n    \\ d\df a - \df \varpi_a \wedge \df e^a
	\end{pmatrix} 
\ee

Applying the projector $\Pi^A{}_I$ to (\ref{firstStructureExtended}), we retrieve the structure equation for the spacetime torsion $\df T^A = D \df e^A$. However, the extended structure equation has one additional component and thus precisely the number of equations necessary to solve uniquely for $\df \omega^{A}{}_B$. The derivative operator is then completely determined in terms of the extended torsion $\df T^I$ and the extended tetrad $\df e^I$. 

If we restrict to flat spacetimes with a Newtonian potential $e^I = \begin{pmatrix} dt & dx^a & - V dt \end{pmatrix}^T$ and torsionless backgrounds $\df T^I = 0$, one may compute the Christoffel symbols (see \cite{GPR_geometry} for the relevant formulae) and make a direct comparison with \cite{MTW-book}. The only non-zero components are
\begin{align}\label{gravity}
	\Gamma^i{}_{00} = \partial^i V.
\end{align}
The ``free-falling" curves are given by
\be
	X^\nu \nabla_\nu X^\mu = 0 \qquad \qquad \implies \qquad \qquad m \ddot x^i = - m \partial^i V ,
\ee
where $X^\mu = \dot x^\mu/(n_\nu \dot x^\nu)$ is the tangent vector to a curve $x^\mu (t)$, precisely the equation of motion for a massive particle moving in the presence of a gravitational field. This justifies our interpretation of $\df a$ as the Newtonian potential and demonstrates the equivalence of our description with standard geometrical treatments of Newtonian gravity in their common domain of application.\\

Note, that the spin connection \(\df\omega^a{}_b\) is invariant under boosts \cite{GPR_geometry}. In $2+1$ dimensions (which is the relevant case for quantum Hall systems), we can construct an abelian \(S0(2)\)-connection as follows. Consider the extended Galilean connection \(\df\omega_{IJ}\) where the indices have been lowered with the non-degenerate metric \(g_{IJ}\)
\be
	\df \omega_{IJ} = 
	\begin{pmatrix}
		0 & - \df \varpi_b & 0 \\
		\df \varpi^a & \df \omega^a{}_b & 0 \\
		0 & 0 & 0
	\end{pmatrix}.
\ee
Since \(n^I \df \omega_{IJ} = n^J \df \omega_{IJ} = 0\) we have a unique \(\hat{\df\omega}_{AB}\) so that \(\df \omega_{IJ} = \Pi^A{}_I \Pi^B{}_J \hat{\df\omega}_{AB}\) where
\be\label{eq:conn-hat}
		\hat{\df\omega}_{AB} = 
	\begin{pmatrix}
		0 & - \df \varpi_b\\
		\df \varpi_a & \df \omega_{ab}
	\end{pmatrix}
\ee
Then the abelian \(SO(2)\)-connection can be defined by
\be\label{eq:abelian-spin-conn}
	\df\omega = \frac{1}{2}n_A \epsilon^{ABC}\hat{\df\omega}_{BC} = \frac{1}{2}\epsilon^{ab}\df\omega_{ab}
\ee
It can be checked that under a \(SO(2)\) rotation by \(\theta\) this transforms as
\be
	\df\omega \rightarrow \df\omega + d\theta
\ee
while it is invariant under boosts. This form of the spin connection will be useful while writing down the Wen-Zee term for the quantum Hall system.

\subsection{Non-relativistic currents and stress-energy}\label{sub:currents}

Given an effective action for the matter fields describing the system, we can define the currents and stress-energy by varying the action with respect to the background geometry and electromagnetic fields. Following \cite{GPR-fluids} we define currents by \footnote{If the action depends explicitly on the Galilean connection \({\df\omega^I}_J\), i.e. the matter fields have spin, then the stress-energy tensor should be ``improved" with contributions from the spin current. We only consider the spinless case in this work and so we do not need these improvements, We defer the analysis of spinful matter and improved stress-energy to an upcoming paper \cite{GPR-improv}.}
\begin{align}\label{eq:var1}
	\delta S_{\rm eff} = \int d^n x | e | \left( -  \tau^\mu{}_I  \delta e^I_\mu  +  j^\mu   \delta A_\mu\right) 
\end{align}
Here $| e | = \text{det}( e^A_\mu)$ is the volume form. The object $\tau^\mu{}_I$ collects response to spatial metric perturbations, temporal metric perturbations and the gravitational field. In a Galilean  theory, these have no independent existence  since they transform amongst each other as indicated by the lower index $I$. However, $\tau^\mu{}_I$ collects them into a single covariant object which we call the stress-energy-momentum tensor. In components
\begin{align}\label{stressEnergyMomComponents}
	\tau^{\mu}{}_I=
	\begin{pmatrix}
		\varepsilon^t & - p_a & - \rho^t \\
		\varepsilon^i & - T^i{}_a & - \rho^i
	\end{pmatrix},
\end{align}
where in terms of the non-covariant parts we have
\begin{align}
	\delta S_{\rm eff} = \int d^n x | e| \left( - \varepsilon^\mu \delta n_\mu - \frac 1 2 T_{ij} \delta h^{ij}  + p_a \delta e^a_t+ \rho^\mu \delta a_\mu + j^\mu \delta A_\mu \right) .
\end{align}
In a spinless theory, the Ward identity for local Galilean transformations fixes $p_a =\rho_a$ (where we have used the frames to convert the indices) and so this is not an independent current \cite{GPR-fluids}.

Consider a family of observers with velocity $u^\mu$ (we shall always assume worldlines are parameterized so that $n_\mu u^\mu =1$). This extra data allows us to define a metric with lowered indices $\overset u h_{\mu \nu}$ by
\be\label{loweredMetric}
	 \overset u h_{\mu \nu} u^\nu = 0 , \qquad \qquad h^{\mu \lambda}  \overset u h_{\lambda \nu} = \overset u P{}^\mu{}_\nu
\ee
where here $\overset u P{}^\mu{}_\nu =  \delta^\mu{}_\nu - u^\mu n_\nu $ acts as a projector orthogonal to both $n_\mu$ and $u^\mu$ 
\be\label{projector}
	n_\mu \overset u P{}^\mu{}_\nu = 0 , \qquad \qquad \overset u P{}^\mu{}_\nu u^\nu = 0.
\ee

 The energy current, stress, and mass current measured by this family are \footnote{The frame-dependent energy current $\varepsilon^\mu = \tau^\mu{}_I u^I$ was originally introduced in \cite{Jensen:2014aia}.}
\be\label{covariantCurrents}
	 \varepsilon^\mu = \tau^\mu{}_I u^I , \qquad \qquad T^{\mu \nu} = -  \overset u P{}^\mu{}_\lambda \tau^\lambda{}_I  \overset u P{}^{\nu I} , \qquad \qquad \rho^\mu = - \tau^\mu{}_I n^I .
\ee
where $ \overset u P{}^{\mu }{}_I = \overset u P {}^\mu{}_\nu \Pi^{\nu}{}_I$. Here $u^I$ is the \emph{unique} extension of $u^A= \begin{pmatrix}1 & u^a \end{pmatrix}^T$ to the extended representation so that $u^A = \Pi^A{}_I u^I$ and $u_I u^I = 0$. Explicitly, we have
\begin{align}
	u^I =
	\begin{pmatrix}
		1 \\ u^a \\ - \frac 1 2 u^2
	\end{pmatrix}
\end{align}
where we have denoted $u^2 = u_a u^a$. Note that the mass current \(\rho^\mu\) is the only component of the stress-energy that can be separated out independently of \(u^\mu\) and thus all observers agree on.

\section{Symmetries of the microscopic theory}\label{sec:symmetries}

	We assume that the microscopic theory of the quantum Hall system is given by a single component, \emph{spinless} Galilean covariant field which is minimally coupled to a background Bargmann spacetime and electromagnetic field. In this case, the electromagnetic and Newtonian potentials only enter the problem through the gauge-covariant derivative $D_\mu = \partial_\mu - i A_\mu - i m a_\mu$, 
(where we have assumed unit charge for convenience). For instance, a spinless Schr\"odinger field with mass $m$ and zero $g$-factor has the action \footnote{A manifestly Galilean invariant description of the Schr\"odinger field on any Bargmann spacetime was given in Sec. 3.2 \cite{GPR_geometry}.}
\begin{align}
	S = \int d^n x \left( i \psi^\dagger D_t \psi - \frac{1}{2m} |D_i \psi |^2 \right) + \mathrm{interactions} .
\end{align}
Apart from Galilean invariance, such microscopic actions are independent of the background Galilean connection \(\df\omega^A{}_B\), and additionally, invariant under the combined \emph{shift symmetry}\footnote{We only consider the spinless case for direct comparison with \cite{Hoyos:2011ez}. The inclusion of spin does not change the logic of our approach but could give rise to additional shift symmetries. A non-zero $g$-factor however may break this symmetry. In GaAs, we have $g \approx - 0.03$, so that this breaking would be very small \cite{Weisbuch:1977rz}.}
\be\label{oneCompSym}
	\df A \rightarrow \df A + m \df \xi,
	\qquad \qquad \df a \rightarrow \df a - \df \xi
\ee
where $\df \xi$ is an arbitrary \(1\)-form. Upon integrating out the microscopic degrees of freedom, these symmetries will be respected by the effective action i.e. the effective action \(S_{\rm Hall}\) has the following functional dependence on the background geometric and electromagnetic fields
\begin{align}
	S_{\rm Hall} = S_{\rm Hall} [ \df e^A , \df A + m \df a ] .\label{eq:generating_sym}
\end{align}
One immediate consequence of the shift symmetry is that the mass and charge currents of the Hall system are simply related by
\begin{align}\label{sameCurrent}
	\rho^\mu = m j^\mu ,
\end{align}
As we will show, since the combination $\df A + m \df a$ is not invariant under Galilean boosts, there are strong restrictions on how we may include it in a boost invariant effective action. These restrictions lead to the corrections found in \cite{Hoyos:2011ez} for the Chern-Simons and Wen-Zee terms. The constitutive relations for the current (\ref{sameCurrent}) are then quite constrained by demanding the symmetry (\ref{oneCompSym}). The rest of the paper is dedicated to showing how one can do this in a manifestly Galilean invariant manner.

\section{The effective drift velocity, gauge field and Galilean connection}\label{sec:frame}

In this section we describe our approach to imposing the shift symmetry (\ref{oneCompSym}) in our effective action. One cannot simply write the standard terms for a gauge field using the combined gauge field $\df A + m \df a$ since this combination is not boost invariant,
\begin{align}\label{gaugeFieldBoost}
	\df A + m \df a \rightarrow \df A + m \lb( \df a + k_a \df e^a - \frac 1 2 k^2\df n \rb) .
\end{align}
i.e. the Chern-Simons and Wen-Zee terms will not be invariant under $Gal(d)$ if we simply perform the replacement $\df A \rightarrow \df A + m \df a$. Viewing the boost transformation of the combined gauge field as a ``modified diffeomorphism", as demonstrated in \cite{Jensen:2014aia}, this is the statement found in \cite{Hoyos:2011ez} that the Chern-Simons and Wen-Zee terms are not diffeomorphism invariant.

Hoyos and Son \cite{Hoyos:2011ez} solved this problem by correcting these terms order by order in an $m$-expansion. With the correction terms included, the action is then boost invariant to the order in $m$ being considered. We take a different approach, defining a single, boost invariant effective gauge field $ \df{\mathcal A}$ by dressing it with appropriate additional contributions from the background geometric fields. Similarly, we define an effective Galilean connection \(\tilde{\df\omega}^A{}_B\) which also respects the shift symmetry. The macroscopic description of the Hall system directly couples only to these effective connections and thus, the effective action we propose (\ref{effectiveAction}) is manifestly Galilean invariant to all orders in $m$. Expanding to first order in \(m\) we reproduce the corrections obtained by \cite{Hoyos:2011ez}. One may now use our formalism to easily construct effective actions to higher order in derivatives, something that would have been very cumbersome in the previous approach.\\

The non-trivial boost transformation (\ref{gaugeFieldBoost}) may be removed in the presence of a preferred velocity $u^\mu$ by simply agreeing to always take $\df a$ to be measured in a frame where \(e^\mu_0 = u^\mu\), denoted as $\overset u {\df{a}}$. Expressed in an arbitrary frame where $u^a \neq 0$, this is
\eq{
 \overset{u}{\df{a}} = \df e^I u_I =  \df a + u_a \df e^a - \frac 1 2 u^2 \df n,
}
where $u^I$ is the null extension of $u^\mu$ discussed in \autoref{sub:currents}. This is invariant under local Galilean transformations. We can then construct 
\begin{align}\label{boostInvA}
	\df \cA = \df A + m \overset{u}{\df{a}},
\end{align}
which is invariant under \eqref{oneCompSym}. $\df \cA$  corresponds to the improved gauge field of Son \cite{Son:2013}, invariant under modified diffeomorphisms, and was stated in this language by Jensen in \cite{Jensen:2014aia}. 

In (\ref{boostInvA}) the velocity \(u^\mu\) is arbitrary, but a quantum Hall system (with non-zero magnetic field) always has a preferred drift velocity
\be\label{EMFrame}
	u^\mu = \frac 1 {2B} \varepsilon^{\mu \nu \lambda} F_{\nu \lambda},
	\quad\text{i.e.}\quad
	u^A = 
	\begin{pmatrix}
		1 \\
		\frac{\epsilon^{ab} E_b}{B}
	\end{pmatrix} .
\ee
Here $B = \frac 1 2 \varepsilon^{\mu \nu \lambda} n_\mu F_{\nu \lambda}$ is the magnetic field, invariant under boosts, while the electric field is $E_a = e^i_a F_{it}$.
One could use this drift velocity to define $\df \cA$ in (\ref{boostInvA}) and proceed to construct boost invariant actions. However, this would violate the shift symmetry (\ref{oneCompSym}) since the drift velocity $u^\mu$ refers only to $\df A$ i.e. it is the drift velocity in an external electromagnetic field. Next we demonstrate how to construct a preferred effective drift velocity in a manner that treats the electromagnetic and Newtonian gravitation fields symmetrically so that (\ref{oneCompSym}) is preserved.\\

To begin, assume there exists a preferred velocity $u^\mu$ constructed from the background geometric data $\df e^I$ and $\df A$, and define the corresponding boost invariant gauge field using this velocity $\df \cA = \df A + m u_I \df e^I$ according to (\ref{boostInvA}). This has a Galilean boost and shift invariant field strength $\df \cF = d \df \cA$. Now, consider the drift velocity \(\mathcal U^\mu\) determined by this invariant field strength
\begin{align}
	\mathcal U^\mu = \frac{1}{2 \mathcal B} \varepsilon^{\mu \nu \lambda} \mathcal F_{\nu \lambda},
	\qquad \qquad \text{where}\qquad \qquad
	\mathcal B = \frac 1 2 \varepsilon^{\mu \nu \lambda} n_\mu \mathcal F_{\nu \lambda} .
\end{align}
This drift velocity feels both electromagnetic and gravitational forces in a symmetric manner. The physical principle behind defining \(\mc U^\mu\) is that a single constituent system (with a single charge to mass ratio) cannot distinguish between electric and gravitational forces or between magnetic and Coriolis forces, and so they should appear together in our treatment. For our construction to be $\df A$ and $m\df a$ symmetric, we insist that \(\mc U^\mu\) is the preferred velocity used to construct $\df \cA$ in the first place i.e. we demand that \(u^\mu\) satisfy the self-consistency equation
\be\label{selfConsistentFrame}
	u^\mu = \mathcal U^\mu = \frac{1}{2 \mathcal B} \varepsilon^{\mu \nu \lambda} \mathcal F_{\nu \lambda} \qquad \qquad \text{with}\qquad \qquad  \df\cF = d(\df A + m\overset{u}{\df a}) .
\ee
 We call a specific solution to this equation \emph{the effective drift velocity}. This effective drift velocity determines a Galilean boost and shift invariant \emph{effective gauge field} \(\df\cA\) from (\ref{boostInvA}). This construction appeared in  \cite{Andreev:2014gia,Moroz:2015jla} in the language of non-relativistic general coordinate invariance and not in the Newton-Cartan formalism.

On flat backgrounds with no gravitational field (\ref{selfConsistentFrame}) becomes
\begin{align}
	m\left(\dot{u}^i +u^j \partial_j u^i\right) = E^i + \epsilon^{ij}u_j B,
\end{align}
which is simply the equation of motion for a charged particle in an external electromagnetic field. In the limit where the electromagnetic field is static and homogeneous this can be solved by the drift velocity. In fact, on a general Bargmann spacetime the consistency equation \eqref{selfConsistentFrame} is equivalent to
\be\label{eq:Hall-fluid}
m u^\nu \nabla_\nu u^\mu = m (\df f + u_a\df T^a - \tfrac{1}{2} u^2 d\df n)^\mu{}_\nu u^\nu + {F^\mu}_\nu u^\nu
\ee
which is the equation of motion for a (unit) charged particle (see Sec.3.1 of \cite{GPR_geometry}). Thus, \(u^\mu\) defined by (\ref{selfConsistentFrame}) is simply the velocity of a congruence of noninteracting particles moving in a curved Bargmann geometry acted upon by electromagnetic, gravitational and torsional forces. \\

 When the electromagnetic field is homogeneous in flat space a solution to (\ref{selfConsistentFrame}) is cyclotron motion about guiding centers at the drift velocity
\begin{align}\label{cyclotron}
	u^i = 
	\begin{pmatrix} 
		\frac{E_y}{B} + \Delta u  \sin \left( \w_c t + \phi \right) \\
		 - \frac {E_x}{B} + \Delta u \cos \left(\w_c t + \phi \right)
	\end{pmatrix},\qquad \qquad \w_c = \frac{B}{m}.
\end{align}
Note that if $\Delta u \neq 0$, the solution is oscillating about the drift velocity at the cyclotron frequency, which is inappropriate for constructing a low-energy effective field theory. We will therefore always choose the solution that is smooth as $m \rightarrow 0$ which necessitates $\Delta u = 0$. This guarantees that our solution will match the electromagnetic drift velocity (\ref{EMFrame}) as $m\rightarrow 0$. Such a solution always exists and we shall refer to it as the \emph{effective drift velocity}. We can then consider the solution in a $m$-expansion (which is also a derivative expansion),
\eq{
u^\mu = u^\mu_{(0)} + m u^\mu_{(1)} + m^2 u^\mu_{(2)} + \ldots
}
Using this we can solve (\ref{selfConsistentFrame}), recursively to any order we need. For instance,
\be\begin{split}
	u^\mu_{(0)} & = \frac{1}{2 B} \varepsilon^{\mu \nu \lambda} F_{\nu \lambda} \\
	u^\mu_{(1)} & = \frac{1}{B}\lb( \half \ce^{\mu\nu\lambda} {\cF_{\nu\lambda}}_{(1)} - \cB_{(1)} u^\mu_{(0)} \rb)
\end{split}\ee
where
\be\begin{split}
	\df\cF_{(1)} & = d(u^I_{(0)}\df e_I) = 2\d_{[\mu} \lb( a_{\nu]} + {\ce_{\nu]}}^\lambda \frac{E_\lambda}{B} - \frac{1}{2}\frac{E^2}{B^2}n_{\nu]} \rb) \\
	\cB_{(1)} & = \ce^{\mu\nu} \d_\mu a_\nu - \frac{1}{4}\frac{E^2}{B^2}\ce^{\mu\nu} (d\df n)_{\mu\nu} - \partial^\mu \left(\frac{E_\mu}{B}\right)
\end{split}\ee
where $E_\mu = F_{\mu\nu} e_0^\nu$. Similarly, at any given order \(O(m^k)\), (\ref{selfConsistentFrame}) is algebraic in \(u^\mu_{(k)}\) (but depends on derivatives of \(u^\mu_{(i<k)}\)) and hence can be solved recursively. In this work we shall only need the \(u^\mu_{(0)}\) solution for comparison with \cite{Hoyos:2011ez} (see \autoref{app:HoyosSon}).\\

We can now write down a Chern-Simons term for a single-constituent Galilean system, we simply write the usual term but for the effective gauge field $\df \cA$. Now we turn to the Wen-Zee term, which was originally \cite{Wen:1992ej} defined to describe response to spatial curvature, and therefore was constructed from the \emph{spatial} frames $e^i_a$. For instance, the torsion-free $SO(2)$-connection for a two-dimensional Riemann surface is simply
\eq{
\w_i = \half g_{ij}\ce^{ab}[e_a,e_b]^j\label{eq:so2_curv}
}
which is defined via a Lie bracket and makes no reference to a specific derivative operator. Recall that, since we assume that the microscopic theory is spinless\footnote{The shift is nonzero even for spinless electrons, as can be seen by quantizing a spinless Schr\"odinger field on a sphere with magnetic flux \cite{Tamm:1931dda}.} our action cannot depend on the background Galilean connection \eqref{eq:generating_sym}, so we must construct one out of our metric data and potentials.\footnote{The generalization of the Wen-Zee term to Lorentzian theories was constructed in \cite{Golkar:2014}.}

	We can specify a Galilean connection uniquely by specifying the extended torsion tensor \(\df T^I\) (\ref{eq:T-decomp}). When the temporal torsion vanishes \(d\df n =0\), we can invariantly set \(\df T^I = 0\) (see Sec. 2.4 \cite{GPR_geometry}), which determines the completely torsion free Newtonian connection. This poses two problems:
\begin{enumerate*}[label=(\arabic*)]
	\item The Newtonian connection does not respect the shift symmetry \eqref{oneCompSym}, as one of the vanishing torsion equations (see (\ref{eq:T-decomp})) is $\df \varpi_a \wedge \df e^a = d\df a$.
	\item If $d\df n \neq 0$ we can no longer set any individual components of the torsion to zero in a boost invariant manner (see Sec. 2.4 \cite{GPR_geometry}).
\end{enumerate*}

 Thus, our strategy is to define an \emph{effective Galilean connection} \(\tilde{\df \omega}^A{}_B\) by specifying a torsion tensor that depends on the effective drift velocity and obeys the shift symmetry by demanding
 \be\label{eq:eff-T}
 \df T^I(\tilde{\df \omega}) = u^I d\df n + n^I d\overset{u}{\df{a}}.
 \ee
This is a manifestly covariant equation and we can inspect components to confirm that all explicit $\df a$ dependence cancels,
\be\label{eq:eff-T-decomp}
\begin{pmatrix}
d\df n \\
d \df e^a + \tilde{\df \omega}^a{}_b \wedge \df e^b + \tilde{\df\varpi}^a \wedge \df n \\
d \df a - \tilde{\df \varpi}_a \wedge \df e^a
\end{pmatrix}
=
\begin{pmatrix}
d\df n \\
u^a d\df n \\
-\tfrac{1}{2}u^2d\df n + d\df a + d( u_a \df e^a) - \half d (u^2\df n)
\end{pmatrix}.
\ee
Thus, our effective Galilean connection \(\tilde{\df\omega}^A{}_B\) depends explicitly only on the frames \(\df e^A\) and the effective drift velocity \(u^\mu\), and consequently satisfies the shift symmetry.

Rewriting (\ref{eq:Hall-fluid}) in terms of the covariant derivative \(\tilde\nabla\) determined by the connection \(\tilde{\df\omega}^A{}_B\) we get
\be
m u^\nu \tilde\nabla_\nu u^\mu = {\cF^\mu}_\nu u^\nu
\ee
i.e. with respect to the derivative \(\tilde\nabla\) the only force on \(u^\mu\) is due to the modified field strength \(\df\cF\). This is simply a manifestation the shift symmetry (\ref{oneCompSym}) of the microscopic description which treats electromagnetic and gravitational fields on an equal footing.

The effective description of the quantum Hall system then only depends on the unique connection \(\tilde{\df\omega}^A{}_B\) determined by \eqref{eq:eff-T}. In everything that follows we will only use this connection, so we drop the ``tilde" notation and denote the connection by $\df \w^A{}_B = \df \w^A{}_B(\df e^A,u^\mu(\df A + m \df a))$. Following (\ref{eq:abelian-spin-conn}) we can define the effective spin connection as 
\begin{align}
	\df \omega = \frac 1 2 \epsilon^{ABC}n_A \hat{\df\omega}_{BC} ,
\end{align}
which is boost invariant and transforms as an $SO(2)\cong U(1)$ connection under local rotations, making it perfect for use in constructing a Wen-Zee term. One may check that it reduces to (\ref{eq:so2_curv}) when considering backgrounds where $u = \d_t$ (i.e. the electric field vanishes) and metric curvature is purely spatial and static. When the electric field is non-vanishing this will reproduce the corrections to Wen-Zee used to restore Galilean invariance in \cite{Hoyos:2011ez}.
\section{The effective action for quantum Hall systems}\label{sec:eff-action}
Using the data defined in the previous section we can now write an effective action \(S_{\rm Hall}\) for the quantum Hall system which has the shift symmetry (\ref{oneCompSym}), in a manifestly Galilean invariant way: the effective action \(S_{\rm Hall}\) is a local and covariant functional of the coframe $\df e^A$, the effective gauge field $\df \cA = \df A + m u_I \df e^I$ and the effective Galilean connection $\df  \omega^A{}_B$.

To do this explicitly, let us select a power counting scheme and proceed with the construction order-by-order. For purposes of comparison, we will adapt the scheme of \cite{Hoyos:2011ez}. The only dimensionful constants are the magnetic length $\ell_B = \sqrt \frac{\hbar c}{|e| B}$ and the cyclotron frequency $\omega_c = \frac{|e| B}{m c}$. Terms are organized as an expansion in a small parameter $\epsilon$ which indicates the strength of perturbations about a flat space background with constant magnetic field. Due to the presence of a preferred time direction, we are free to select the strength of temporal and spatial components of perturbations independently
\begin{align}
	\partial_t \sim \epsilon^2 \omega_c,
	&&\partial_i \sim \epsilon \ell_B^{-1},
	&&\delta e^A_\mu \sim 1,
	&&\delta \mathcal A_t \sim \epsilon^0 \omega_c,
	&&\delta \mathcal A_i \sim \epsilon^{-1} \ell_B^{-1}.
\end{align}
In this counting, the magnetic field may have \(O(1)\) variations $\delta \mathcal B \sim \epsilon^0 \ell_B^{-2}$ (so long as the gap doesn't close) but the electric field is small $\mathcal E_i \sim \epsilon \omega_c \ell_B^{-1}$. We can easily compute the power counting of the spin connection
\begin{align}
	 \omega_t \sim \epsilon^2 \omega_c,
	&& \omega_i \sim \epsilon \ell_B^{-1}
\end{align}
so that the curvature $\df R = d\df\omega$ is order $\epsilon^2$: $\delta  \df R \sim \epsilon^2 \ell_B^{-2}$.

To second order in $\epsilon$, the most general action we can write is \footnote{\(\epsilon(\cB)\) is the equation of state term which is order \(0\) in power counting; we hope the reader forgives this dual use of \(\epsilon\). Terms such as $n d\cA$ and $udu$ are at lowest order proportional to $\cB$ and are not independent of the \(\epsilon(\cB)\) term.}
\begin{subequations}\label{effectiveAction}\begin{align}
	S_{\rm Hall} & = \int d^3 x | e | \sum \mathcal L_i \nonumber \\
	&\text{where} \nonumber \\
	\cL_{-1} &= \frac{\nu}{4\pi} \ce^{\mu\nu\rho}\cA_\mu \d_\nu \cA_\rho , \label{eq:eff-CS}\\
	\cL_0 &= -\e(\cB) , \label{eq:eff-eos}\\
	\cL_1 &=\frac{\k}{2\pi} \ce^{\mu\nu\rho}\w_\mu \d_\nu \cA_\rho + f G , \label{eq:eff-WZ}\\
	\cL_2 &= g_1  R + g_2 u^\mu \nabla_\mu \mathcal B + g_3 \nabla_\mu\cB\nabla^\mu\cB +g_4  G^2 + g_5 G^\mu \nabla_\mu \cB .\label{eq:eff-curv}
\end{align}\end{subequations}
Here \(R = \ce^{\mu\nu}R_{\mu\nu} = 2\ce^{\mu\nu\rho}n_\mu \d_\nu\w_\rho \) is the spatial Ricci scalar of the effective spin connection, and $f, g_i$ are arbitrary functions of $\mathcal B$. We have also defined the quantities
\be
	G^\mu = - \half \ce^{\mu\nu\rho} (d\df n)_{\nu\rho} , \qquad \qquad  G = -  n_\mu G^\mu .
\ee
Note that \(G\) will vanish on any causal Bargmann geometry (\(\df n \wedge d\df n =0 \)) and \(G^\mu\) will vanish on any geometry with absolute time (\(d\df n = 0\)) \cite{GPR_geometry}, but it is necessary to consider an unrestricted \(\df n\) to fully capture energy transport.

\subsection{Comparison with Hoyos and Son}\label{retHS}
Let us compare the effective action (\ref{effectiveAction}) with the improvement terms of Hoyos and Son \cite{Hoyos:2011ez}. Recall that in \cite{Hoyos:2011ez}, Galilean invariance is imposed by demanding invariance under ``time dependent spatial diffeomorphisms" generated by a spatial vector field $\xi^i$, where the gauge field transforms ``anomalously" under infinitesimal diffeomorphisms as
\begin{align}\label{modifiedDiff}
	\delta A_0 = - \xi^k \partial_k A_0 - A_k \dot \xi^k ,
	&&\delta A_i = - \xi^k \partial_k A_i - A_k \partial_i \xi^k - m h_{ik}  \dot \xi^k .
\end{align}

Under this transformation, the Chern-Simons, Wen-Zee and equation of state terms are no longer invariant. However, \cite{Hoyos:2011ez} impose diffeomorphism invariance by hand, by including correction terms that cancel the modified transformation at every order in \(m\). This yields (up to \(O(m)\))
\begin{align}\label{HSimprovement}
	&\mathcal L_{\text{CS}} = \frac{\nu}{2 \pi} \left( \varepsilon^{\mu \nu \lambda} A_\mu \partial_\nu A_\lambda + \frac m B h^{ij} E_i E_j \right) ,\nonumber \\
	&\mathcal L_\text{WZ} = \frac{\kappa}{2 \pi} \left( \varepsilon^{\mu \nu \lambda} ( \omega_\text{HS} )_\mu \partial_\nu A_\lambda + \frac{1}{2 B} h^{ij} \partial_i B E_j \right), \nonumber \\
	&\mathcal L_\text{EOS} = - \epsilon (B) - \frac m B \epsilon'' (B) h^{ij} \partial_i B E_j .
\end{align}
where \(\df \omega_\text{HS}\) is the choice of spin connection used by Hoyos and Son \cite{Hoyos:2011ez} given by
\begin{align}\label{eq:HS-conn}
	(\omega_\text{HS})_t = \frac 1 2 \epsilon_{ab} e^{aj} \partial_t e^b_j,
	&&(\omega_\text{HS})_i = \frac 1 2 \epsilon_{ab} e^{aj} \nabla^{(s)}_i e^b_j
\end{align}
 with $\nabla^{(s)}$ being the spatial derivative operator on a constant time slice.

In this picture, the $O ( m^1 )$ transformation of the first terms exactly cancel the $O (m^1)$ part of the transformation of the second. There is an $O(m^2)$ part that remains uncanceled, but if the diffeomorphisms under consideration are not too large, we have invariance to the order being considered (specifically, they must scale as $\xi^i \sim O ( \epsilon^{-2})$). In principle one can continue to correct these terms order by order in $m$ in the above way, obtaining an infinite series in $m$ that is totally invariant.

Comparing (\ref{modifiedDiff}) with (\ref{gaugeFieldBoost}), we see that the modified diffeomorphism transformation reflects the fact that the gauge field used in \cite{Hoyos:2011ez} was \(\df A + m\df a\) which is not boost invariant. The correction terms obtained by \cite{Hoyos:2011ez} compensate for the boost transformation order-by-order in \(m\).\\

On the other hand, since we use the manifestly boost invariant effective gauge field determined by the effective drift velocity \(u^\mu\), the action (\ref{effectiveAction}) has implicit dependence on $m$ that extends to arbitrarily high order, and is manifestly Galilean and shift invariant. We check our approach for consistency with Hoyos and Son by expanding (\ref{effectiveAction}) in $m$. For details of this computation we refer the reader to  \autoref{app:HoyosSon} and simply present the final results here.

\begin{subequations}\begin{align}
	\mathcal L_\text{CS} & = \frac{\nu}{4\pi} \left( \ce^{\mu\nu\rho}A_\mu \d_\nu A_\rho +  \frac{m}{B} E^2 + 2 m~ \ce^{\mu\nu\rho}a_\mu \d_\nu A_\rho \right) + \cdots \\
	\mathcal L_\text{EOS} & = - \e(B) - \frac{m}{B} \e''(B)\partial^\mu B E_\mu - m \e'(B)\left( \ce^{\mu\nu}\d_\mu a_\nu - \frac{1}{4}\frac{E^2}{B^2}\ce^{\mu\nu} (d\df n)_{\mu\nu} \right) + \cdots \\
	\mathcal L_\text{WZ} & = \frac{\k}{2\pi} \lb( \ce^{\mu\nu\rho}(\w_{\rm (HS)})_{\mu} \d_\nu A_\rho + \frac{1}{2} \d_\mu B \frac{E^\mu}{B} + \frac{1}{2}B \ce_{\mu\nu} \partial^\mu e_0^\nu \rb) + \cdots
\end{align}\end{subequations}
where $E^\mu$ is the boost non-invariant electric field measured in our Galilean frame, emphasizing that this expansion costs us manifest Galilean invariance.

We note that, we have computed the corrections to \(\cL_{\rm CS}\) and \(\cL_{\rm EOS}\) up to \(O(m)\) on any Bargmann spacetime. The corrections to \(\cL_{\rm WZ}\) are computed on a spacetime satisfying \(d\df n =0\) at \(O(m^0)\) and arise from the fact that the effective spin connection \(\df\omega\) we use differs from the connection \(\df\omega_{\rm HS}\) used by Hoyos and Son (see \autoref{app:HoyosSon} for details) when there is a non-vanishing electric field. The \(G\)-dependent terms in (\ref{effectiveAction}) were not considered in \cite{Hoyos:2011ez} since they vanish when \(d\df n = 0\), and the term with time derivative of \(\cB\) is absent since they assume a static magnetic field. The gravitational correction to $\mathcal L_\text{CS}$ has a particularly simple physical interpretation. In a frame where $\df a = - V dt$ the correction is $- m \frac{\nu B}{2 \pi} V$, the gravitational potential energy of a mass density $\rho^0 = m\frac{\nu B}{2 \pi}$; in (\ref{eq:1pt-func}) we compute that this is indeed the effective mass density of the quantum Hall system. The other extra terms arise due to Coriolis forces and torsion in the background Bargmann spacetime.\\

We thus see that the effective action (\ref{effectiveAction}) reproduces the results of Hoyos and Son in their common domain of applicability, where the temporal torsion and gravitational field vanish, and the magnetic field is static. Since our objective is to compare our effective action with that of \cite{Hoyos:2011ez}, we do not compute the \(O(m)\) corrections to (\ref{eq:eff-WZ}) and (\ref{eq:eff-curv}). We leave this, and the task of computing higher order \(m\)-corrections to the enterprising reader. We emphasize again the effective action (\ref{effectiveAction}) is both Galilean and shift invariant at \emph{any} order in \(m\).

\section{Linear response}\label{sec:lin-response}

%
As we are using an effective field theory description we can only trust results in non-dissipative channels. We will therefore only present non-dissipative linear response results. We shall consider linear response about a flat background (in an inertial frame) with a magnetic field \(B\) and vanishing electric and gravitational field.
\be\label{trivialBackground}
	\df e^I = 
	\begin{pmatrix} 
		dt \\
		d x^a \\
		0
	\end{pmatrix},
	\qquad \df T^I = 0 ,\qquad \df  A = B x dy , \qquad u^\mu = (\d_t)^\mu.
\ee

Note that on the trivial background the coordinate frame \(\d_t\) and effective drift velocity \(u^\mu\) coincide, so there is no difference between the components of $\tau^\mu{}_I$ in equation (\ref{stressEnergyMomComponents}) and the covariant currents (\ref{covariantCurrents}).

If we consider one-point functions on this background using \eqref{eq:var1} we find
\eq{
\delta S = \int\left[
\frac{\nu B}{2\pi} (\delta A_t + m \delta a_t) - \e(B) \delta n_t  + (B \e'(B) - \e(B)) (\delta e^1_x + \delta e^2_y)
\right],
}
which gives charge \(j^0\), mass \(\rho^0\), energy density \(\ce^0\) and pressure \(P\)
\eq{\label{eq:1pt-func}
j^0 = \nu N_\phi , \qquad  \rho^0 = m j^0 , \qquad  \ce^0 = \e(B) , \qquad P = B \e'(B) - \e(B),
}
confirming that $\e(B)$ is the equation of state for the incompressible quantum Hall fluid.

Calculating two-point functions is similarly straightforward. The action generates the retarded two point functions
\begin{align}
	&G^{\mu,\nu}_{jj} ( x - y ) = \frac{\delta^2 S}{\delta A_\nu (y) \delta A_\mu (x)} = \left \langle \frac{\delta j^\mu (x) }{\delta A_\nu (y)} \right \rangle + i \theta (x^0 - y^0 ) \left \langle [ j^\mu (x) , j^\nu ( y) ] \right \rangle ,\nonumber \\
	&G^{\mu,\nu}_{j \varepsilon} ( x - y ) =  \frac{\delta^2 S}{\delta n_\nu (y) \delta A_\mu (x)} = \left \langle \frac{\delta j^\mu (x) }{\delta n_\nu (y)} \right \rangle - i \theta (x^0 - y^0 ) \left \langle [ j^\mu (x) , \varepsilon^\nu ( y) ] \right \rangle ,\nonumber \\
	&G^{\mu,ij}_{j T} ( x - y ) =  \frac{\delta^2 S}{\delta h_{ij} (y) \delta A_\mu (x)} = \left \langle \frac{\delta j^\mu (x) }{\delta h_{ij} (y)} \right \rangle + \frac i 2 \theta (x^0 - y^0 ) \left \langle [ j^\mu (x) , T^{ij} ( y) ] \right \rangle ,\nonumber \\
	&G^{\mu,\nu}_{\varepsilon \varepsilon} ( x - y ) =-  \frac{\delta^2 S}{\delta n_\nu (y) \delta n_\mu (x)} = \left \langle \frac{\delta \varepsilon^\mu (x) }{\delta n_\nu (y)} \right \rangle - i \theta (x^0 - y^0 ) \left \langle [ \varepsilon^\mu (x) , \varepsilon^\nu ( y) ] \right \rangle ,\nonumber \\
	&G^{\mu,ij}_{\varepsilon T} ( x - y ) = -\frac{\delta^2 S}{\delta h_{ij} (y) \delta n_\mu (x)} = \left \langle \frac{\delta \varepsilon^\mu (x) }{\delta h_{ij} (y)} \right \rangle + \frac i 2 \theta (x^0 - y^0 ) \left \langle [ \varepsilon^\mu (x) , T^{ij} ( y) ] \right \rangle ,\nonumber \\
	&G^{ij,kl}_{TT} ( x - y ) = 2\frac{\delta^2 S}{\delta h_{kl} (y) \delta h_{ij} (x)} = \left \langle \frac{\delta T^{ij} (x) }{\delta h_{kl} (y)} \right \rangle + \frac i 2 \theta (x^0 - y^0 ) \left \langle [ T^{ij} (x) , T^{kl} ( y) ] \right \rangle ,
\end{align}
which we will present in Fourier space
\begin{align}
	G ( k ) &= \int d^n x e^{i ( \omega_\epsilon x^0 - k_i x^i )} G (x),
\end{align}
where $\omega_\epsilon = \omega + i \epsilon$ for small $\epsilon>0$.

 The only complication in evaluating these that is not present in standard calculations is the dependence of the action on the self-consistent frame at linear order. We then need to solve (\ref{selfConsistentFrame}) for the effective drift velocity perturbatively about the background (\ref{trivialBackground}). This is easily done in Fourier space,
\eq{
\delta u^i(\w,k) = \frac{i B \ce^{ij} (\w \a_j + \a_0 k_j) + B^2 e^i_t + m \w (k_i \a_t + \w a_i + i B \ce^{ij} e_t^j)}{B^2 - m^2\w^2}, 
}
using the shorthand $ \df \alpha = \df A + m\df  a$. 

%

We first compute the transport coefficients that reproduce results already known in the literature as a consistency check. Then, we present new results on energy transport in the quantum Hall system. It is a straightforward computation to work out the electromagnetic perturbations of the effective action \(S_{\rm Hall}\) \eqref{effectiveAction} about the background \eqref{trivialBackground}. A quick calculation demonstrates that the Hall conductivity is
\eq{\label{eq:hall_cond_k2}
\sigma_{xy} = \frac{\nu}{2\pi} + \left(\frac{\k}{4\pi B} - \frac{m \e''(B)}{B}\right)k^2 +O(k^3).
}
We can also measure $\e''(B)$ independently by turning on inhomogeneous, static perturbations \(\delta B\) in the magnetic field, where we find
\eq{\label{eq:current_inhom_B}
\vev{j^i} = - \e''(B) \ce^{ij} \d_j \delta B+\cO(k^3),
}
reproducing the main results of \cite{Hoyos:2011ez}.\footnote{ The results \eqref{eq:hall_cond_k2} and \eqref{eq:current_inhom_B} were obtained from a slighly different method in \cite{Son:2013}.}

The parity-odd current response of the system to a homogeneous electric field \(E_i\) is determined entirely by the Chern-Simons term (\ref{eq:eff-CS}) for the effective gauge field (as guaranteed by the Kohn-Luttinger theorem \cite{Kohn:1961zz}),\footnote{Note that here we have not taken the explicit derivative expansion, and we find that we can reproduce the full nonlinear frequency dependence derived from \eqref{trivialBackground}. Gauge invariance guarantees that no other terms in our effective action can contribute to current response to a homogeneous electric field.}
\eq{
\sigma_{xy}(\w,0) =\frac{\nu}{2\pi}\frac{B^2}{B^2-m^2\w^2}},
i.e.
\eq{\vev{j^i(\w,0)} = \frac{\nu }{2\pi}\frac{B^2}{B^2 - m^2 \w^2}  \ce^{ij}E_j(\w,0).}
In real space this gives
\eq{
\vev{j^i(t)} = \frac{\nu \w_c}{2\pi } \int dt' \Theta(t-t') \sin\!\left[\w_c(t-t')\right] \ce^{ij}E_j(t').
}
We can also calculate the static susceptibility,
\eq{
\chi(q)=-\vev{j^0(0,k)j^0(0,-k)}= - \frac{\nu m}{2\pi B}k^2 + \cO(k^3).
}
Both of these results match the calculations of \cite{Son:2013}.

We can also calculate the Hall viscosity \(\eta_H\), the parity-odd component of the viscosity tensor. We find
\eq{
\vev{\tau^{xx}(\w,0) \tau^{xy}(-\w,0)} = i \eta_H \w+\cO(k^3) \qquad \qquad\text{with}\qquad \qquad \eta_H = \frac{\kappa B}{4\pi},
}
as expected. We can also see the Hall viscosity in the stress response to electric perturbations,
\eq{
\vev{\tau^{ij}} = - \frac{\kappa}{2\pi} \d^{(i}E^{j)}.
}

The charge current response to a spatial metric perturbation \(h^{ij}\) is
\begin{subequations}\begin{align}
\delta\vev{j^0} &= \frac{\k}{4\pi} \left(R(h) -\half  \d^i \d_i h \right)+ \frac{m \e''(B)}{2} \d^i \d_i h, \label{eq:current_curv}\\
\vev{j^i} &= -\frac{B \e''(B)}{2} \ce^{ij}\d_j h.
\end{align}\end{subequations}
where \(h\) is the trace of the metric perturbation (which corresponds to an expansion of the system) and \(R(h)\) is the spatial Ricci scalar. Noting that all but the first term in \eqref{eq:current_curv} are total derivatives,\footnote{Of course $\df R = d \df \w$ but $\df \w$ is not gauge invariant, yielding the Gauss-Bonnet theorem.} using the Gau\ss-Bonnet theorem (along with the one-point function in (\ref{eq:1pt-func})) we get 
\eq{
Q(\Sigma_g) = \nu N_\phi + 2\kappa(1-g), \qquad \qquad \kappa = \half \nu \cS.
}
which reproduces the shift in the charge of the Hall system found by \cite{Wen:1992ej}.\\

Since, the effective action (\ref{effectiveAction}) couples in a Galilean invariant manner to a general Bargmann geometry (\(\df n\) is unrestricted), we can explicitly compute transport induced by varying \(\df n\). We note that our definition of energy response is the same as that outlined in for instance \cite{Geracie:2014nka} (which does not explicitly construct transport but constructs the energy density and current operators for a microscopic theory).  Consider turning on a small inhomogeneous Luttinger potential \(\Phi_L\) through $\df n = e^{-\Phi_L} dt$, we find the parity-odd current and energy response is
%
%
%
%
\eq{
\vev{j^i} = - \e'(B) \ce^{ij} \d_j e^{-\Phi_L},\qquad \qquad \vev{\ce^i} = f(B) \ce^{ij} \d_j  e^{-\Phi_L}.
}
that is the induced energy and charge currents  are pure curl. These results are consistent with the energy transport calculations done in \cite{Geracie:2014zha} (which projected to the lowest Landau level) and \cite{Bradlyn:2014wla}, which considered systems with rotational symmetry but not Galilean boosts.

Finally, we can similarly calculate energy currents in response to stress and strain, written most simply in terms of the perturbed spatial metric $h^{ij}$\,
\begin{subequations}\begin{align}
\delta\vev{\ce^0} &= \frac{g_1(B)}{2} R(h) - \half P h,~\\
\vev{\ce^i} &= \frac{B f'(B)}{4} \ce^{ij}\d_j h,
\end{align}\end{subequations}
where \(P\) is the pressure of the Hall fluid calculated in (\ref{eq:1pt-func}). The pressure term reflects the system's response to expansion/compression exactly as expected, and the $g_1$ term allows for energy accumulation at curvature, as the term is (up to normalization) simply $\df n \wedge \df R$. Unfortunately as $\df n$ is not a connection there is no topological quantization of the coefficient. 


\section{Conclusions}\label{sec:the_end}

Starting from the assumption that the microscopic description of a quantum Hall system is a single constituent and spinless field, we have constructed a manifestly Galilean invariant effective action which describes an incompressible quantum Hall state, providing an explanation of the results found in \cite{Hoyos:2011ez} in a manifestly diffeomorphism and Galilean invariant manner. The assumption of minimal coupling to electromagnetic and gravitational potentials along with Galilean invariance strongly constrains the form of the effective action, where all terms are constructed out of an effective drift velocity, gauge field and Galilean connection. The construction allows us to couple the quantum Hall system to any non-relativistic spacetime (i.e. a Bargmann spacetime) and as a result obtain energy and stress transport coefficients. We also explicitly show that our construction reduces to that of Hoyos and Son \cite{Hoyos:2011ez} to leading order in the mass of the microsopic field.

We constructed the effective drift velocity \(u^\mu\) to maintain the shift symmetry (\ref{oneCompSym}) of the microscopic description. This construction is not unique, but other terms one could add to the construction are at higher order in derivatives. Adding higher derivative corrections will simply be equivalent to a redefinition of some higher derivative terms in the effective action. One might wonder whether we could equivalently construct an action by integrating out an arbitrary time-like vector field. We suspect that at lowest order the action is a Lagrange multiplier constraining the arbitrary vector field to be the drift velocity, and massive fluctuations will correspond to higher derivative terms in the action.\\

It would be interesting to construct more general systems, for instance with the microscopic description being fermions directly coupled to a background Galilean connection. This should be a simple generalization of the work presented here which we leave to future work. While inhomogeneous electromagnetic response may be realized in experiments, it would also be interesting to consider experimental methods to measure nontrivial geometric and gravitational response such as non-inertial effects. In natural units the mass-to-charge ratio is $m_e/q_e\sqrt{c \hbar \e_0 }  = 3\times 10^{-30}$, but it is possible that materials with anomalously large band masses and small permittivity may make these effects measurable.

\section*{Acknowledgments}
We would like to thank Dam T. Son for insightful discussions throughout the course of this work and Carlos Hoyos for helpful comments on an early draft. M.G. is supported in part by NSF grant DMR-MRSEC 1420709. K.P. is supported in part by NSF grants PHY~12-02718 and PHY~15-05124. M.M.R is supported in part by DOE grant DE-FG02-13ER41958.

\appendix

\section{Computation of $m$-corrections to the effective action}\label{app:HoyosSon}

To compare our effective action (\ref{effectiveAction}) to the one found by Hoyos and Son \cite{Hoyos:2011ez}, we expand the effective drift velocity \(u^\mu\) defined by (\ref{selfConsistentFrame}) in \(m\) as
\eq{
u^\mu = u^\mu_{(0)} + m u^\mu_{(1)} + m^2 u^\mu_{(2)} + \ldots
}
and solve (\ref{selfConsistentFrame}) order-by-order in \(m\). The extended \(u^I\) then has the corresponding expansion with
\be
	u^I_{(0)} = \begin{pmatrix}1 \\ u^a_{(0)} \\ -\tfrac{1}{2} u^2_{(0)} \end{pmatrix} , \qquad  	u^I_{(1)} = \begin{pmatrix}0 \\ u^a_{(1)} \\ - u^a_{(0)} {u_a}_{(1)} \end{pmatrix} , \qquad 	u^I_{(2)} = \begin{pmatrix}0 \\ u^a_{(2)} \\ -\tfrac{1}{2} u^2_{(1)} - u^a_{(0)} {u_a}_{(2)}. \end{pmatrix}
\ee

	The zeroth order \(u^\mu_{(0)}\) is just the electromagnetic drift velocity
\be\label{eq:u0}
	u^\mu_{(0)} = \frac{\ce^{\mu\nu\rho}F_{\nu\rho}}{2B} , \qquad u^a_{(0)} = \frac{\ce^{ab}E_b}{B} , \qquad u^2_{(0)} = \frac{E^2}{B^2}
\ee
where $E_a = F_{a0}$ is the electric field measured in our frame. The effective gauge field \(\df\cA\) in (\ref{boostInvA}) has the corresponding expansion with
\be
	\df\cA_{(0)} = \df A , \qquad  \df\cA_{(k)} = u^I_{(k-1)}\df e_I \quad\text{for}\quad k \geq 1
\ee
in particular we will need the \(O(m)\) correction which can be computed to be
\be\label{eq:eff-A-1}
	\cA_{\mu(1)} = u^I_{(0)}\df e_I = a_\mu + {\ce_\mu}^\nu \frac{E_\nu}{B} - \frac{1}{2}\frac{E^2}{B^2}n_\mu
\ee

Using this we can compute the \(O(m)\) part of the Chern-Simons term (\ref{eq:eff-CS}) for the effective gauge field as
\be\label{eq:eff-CS-exp}
	\cL_{-1} = \frac{\nu}{4\pi} \left( \ce^{\mu\nu\rho}A_\mu \d_\nu A_\rho +  \frac{m}{B} E^2 + 2 m~ \ce^{\mu\nu\rho}a_\mu \d_\nu A_\rho \right)
\ee
The first \(O(1)\)-term is the Chern-Simons term for electromagnetic field, the second term is the correction obtained by \cite{Hoyos:2011ez} while the third term is a mixed-Chern-Simons term between the mass gauge field and electromagnetic fields (this term vanishes in the absence of gravity).\\

Using (\ref{eq:eff-A-1}) we can compute the \(O(m)\) correction to the effective magnetic field as
\be\label{eq:eff-B-1}
	\cB_{(1)} = \ce^{\mu\nu} \d_\mu a_\nu - \frac{1}{4}\frac{E^2}{B^2}\ce^{\mu\nu} (d\df n)_{\mu\nu} - \partial^\mu \left(\frac{E_\mu}{B}\right)
\ee
and thus any function of \(\cB\) in (\ref{effectiveAction}) can be expanded as \(f(\cB) = f(B) + m \cB_{(1)} f'(B)\), in particular the equation of state term (\ref{eq:eff-eos}) is
\be\label{eq:eff-eos-exp}\begin{split}
	\cL_0 = - \e(\cB) & = - \e(B) + m \e'(B)\partial^\mu \left(\frac{E_\mu}{B}\right) - m \e'(B)\left( \ce^{\mu\nu}\d_\mu a_\nu - \frac{1}{4}\frac{E^2}{B^2}\ce^{\mu\nu} (d\df n)_{\mu\nu} \right) \\
	& = - \e(B) - \frac{m}{B} \e''(B)\partial^\mu B E_\mu - m \e'(B)\left( \ce^{\mu\nu}\d_\mu a_\nu - \frac{1}{4}\frac{E^2}{B^2}\ce^{\mu\nu} (d\df n)_{\mu\nu} \right)
\end{split}\ee
where in the second line we have integrated by parts and discarded the boundary term. Again, the second term is the correction term obtained by \cite{Hoyos:2011ez} and the third term comes from non-trivial background torsion and gravitational field.\\

To get corrections to the effective Wen-Zee term in (\ref{eq:eff-WZ}) obtained by \cite{Hoyos:2011ez}, we need to compare our effective spin connection \(\df\omega\) to the spin connection \(\df\omega_{\rm HS}\) given by (\ref{eq:HS-conn}). For the sake of direct comparison with \cite{Hoyos:2011ez}, we restrict to the case \(d\df n = 0\). We then must solve the equation (\ref{eq:eff-T}) for the effective connection. Writing in components using (\ref{eq:eff-T-decomp}) we solve
\be\begin{split}
d \df e^a + {\df \omega}^a{}_b \wedge \df e^b + {\df\varpi}^a \wedge \df n & = 0 \\
- {\df \varpi}_a \wedge \df e^a & = d\lb( u_a \df e^a - \half u^2\df n \rb) 
\end{split}\ee
Solving for \(\df\omega = \tfrac{1}{2}\epsilon^{ab}\df\omega_{ab}\) gives
\be\begin{split}
	\df\omega & = \df\omega_{\rm HS} + \lb[ \frac{1}{4}\epsilon^{\mu\nu} d ( u_a \df e^a - \tfrac{1}{2} u^2 \df n )_{\mu\nu} + \frac{1}{2}\epsilon_{\mu\nu} \partial^\mu e^\nu_0  \rb] \df n \\
	& = \df\omega_{\rm HS} + \lb[ - \frac{1}{2}\d_\mu \lb( \frac{E^\mu}{B} \rb) + \frac{1}{2}\epsilon_{\mu\nu} \partial^\mu e^\nu_0  \rb] \df n
\end{split}\ee
where in the second line we have used \(d\df n = 0\) and \(u^\mu = u^\mu_{(0)}\) from (\ref{eq:u0}) to compute the leading order corrections. Using this in the effective Wen-Zee term (\ref{eq:eff-WZ}) (and integrating by parts, neglecting the boundary terms) we get at \(O(m^0)\)
\be\label{eq:eff-WZ-0}
	\cL_{WZ} = \frac{\k}{2\pi} \lb( \ce^{\mu\nu\rho}(\w_{\rm HS})_{\mu} \d_\nu A_\rho + \frac{1}{2} \d_\mu B \frac{E^\mu}{B} + \frac{1}{2}B \epsilon_{\mu\nu} \partial^\mu e^\nu_0 \rb) 
\ee
where the second term is precisely the correction found by \cite{Hoyos:2011ez} and the third term arises if the choice of frame \(e^\mu_0 \) is non-inertial. Note that since the effective spin connection differs from the one used by Hoyos and Son only in the ``time component" there are no corrections to the Ricci scalar term in (\ref{eq:eff-curv}) in agreement with \cite{Hoyos:2011ez}.

%

\bibliographystyle{JHEP}
\bibliography{WenZeerefs}

\end{document}